\documentstyle[12pt,aaspp4,epsfig]{article}
\def\deg{\ifmmode^\circ\else$^\circ$\fi}

\def\cf{{\it cf.}}

\def\et{{et al.~}}

\def\hf{\mbox{$H_{1.6}$}}
\def\arcs{\ifmmode {''}\else $''$\fi}
\def\arcm{\ifmmode {'}\else $'$\fi}
\def\parcs{\sa=.07em \sb=.03em
     \ifmmode $\rlap{.}$^{\scriptscriptstyle\prime\kern -\sb\prime}$\kern -\sa$
     \else \rlap{.}$^{\scriptscriptstyle\prime\kern -\sb\prime}$\kern -\sa\fi}
\def\parcm{\sa=.08em \sb=.03em
     \ifmmode $\rlap{.}\kern\sa$^{\scriptscriptstyle\prime}$\kern-\sb$
     \else \rlap{.}\kern\sa$^{\scriptscriptstyle\prime}$\kern-\sb\fi}

\def\spose#1{\hbox to 0pt{#1\hss}}
\def\simlt{\mathrel{\spose{\lower 3pt\hbox{$\mathchar"218$}}
     \raise 2.0pt\hbox{$\mathchar"13C$}}}
\def\simgt{\mathrel{\spose{\lower 3pt\hbox{$\mathchar"218$}}
     \raise 2.0pt\hbox{$\mathchar"13E$}}}
\def\lsim{\rlap{$<$}{\lower 1.0ex\hbox{$\sim$}}}
\def\gsim{\rlap{$>$}{\lower 1.0ex\hbox{$\sim$}}}

\begin{document}

\title{Galaxy Morphology from NICMOS Parallel Imaging\footnote[1]{Based
on observations made with the NASA/ESA {\em Hubble Space Telescope},
obtained from the data archive at the Space Telescope Science
Institute, which is operated by the Association of Universities
for Research in Astronomy, Inc., under NASA contract NAS 5-26555.}}

\author{Harry I. Teplitz\footnote[2]{NOAO Research Associate}, Jonathan
P. Gardner$^2$, Eliot M. Malumuth\footnote[3]{Raytheon STX Corp., Lanham
MD 20706}, Sara R. Heap}

\affil{Laboratory for Astronomy and Solar Physics, Code 681, Goddard
Space Flight Center, Greenbelt MD 20771 \\Electronic mail:
hit@binary.gsfc.nasa.gov, gardner@harmony.gsfc.nasa.gov,
eliot@barada.gsfc.nasa.gov, hrsheap@stars.gsfc.nasa.gov} 

\begin{abstract}
 
We present high resolution NICMOS images of random fields obtained
in parallel to other HST observations. We present galaxy number
counts reaching $H_{1.6}=24$. The $H_{1.6}-$band galaxy counts show
good agreement with the deepest $I-$ and $K-$band counts obtained
from ground-based data. We present the distribution of galaxies
with morphological type to $H_{1.6}<23$. We find relatively fewer
irregular galaxies compared to an $I-$band sample from the Hubble
Deep Field, which we attribute to their blue color, rather than to
morphological K-corrections. We conclude that the irregulars are
intrinsically faint blue galaxies at z$<$1.

\end{abstract}

\keywords{
cosmology: observations ---
galaxies: evolution ---
galaxies: fundamental parameters ---
galaxies: irregular ---
infrared: galaxies ---
surveys
}

\section{Introduction}

The Near-Infrared Camera and Multiobject Spectrometer (NICMOS;
Thompson \et 1998) has taken multi-color images and grism spectra of
random fields in the NICMOS Parallel Program (\cf~Holfeltz \et 1998).
Like the Medium Deep Survey (MDS; Griffiths \et 1994) and the Hubble
Deep Field (HDF; Williams \et 1996), the NICMOS parallel imaging
provides high-resolution images of galaxies at the faintest magnitudes
observed. 

Our understanding of the evolution of galaxies has largely depended on
our knowledge of their place in the Hubble sequence. The MDS provided
the first measure of faint galaxy number-magnitude counts as a
function of morphological type (Driver \et 1995, Glazebrook \et 1995).
It showed a population of rapidly evolving irregular/peculiar galaxies
that appear to explain the faint blue galaxy excess.  At large
redshifts, it becomes problematic to deduce a galaxy's morphology from
optical images, as the observation becomes increasingly dominated by
blue and UV light from young stars, leading to a ``morphological
K-correction'' (O'Connell 1997).  With NICMOS deep imaging, it is now
possible to sample the older stellar population and measure unbiased
galaxy morphologies. 

NICMOS parallel imaging also provides galaxy number counts, as deep or
deeper than are possible from the ground. Both optical and infrared
number count surveys have detected excesses of faint galaxies over
simple no-evolution model predictions (Tyson 1988, Gardner \et 1993;
see Ellis 1997 for a review). Theoretical models can explain the
continued rise of the K-band counts, seen in the recent surveys from
the Keck telescope, using either evolutionary effects or a dwarf-rich
luminosity function -- that is, one with a steep faint-end slope (c.f.
Bershady et al. 1998).  However, ground-based counts at the K=23 level
are based on only $\sim 5$\sq\arcm~total over 6-7 fields of view, so
the critical last few magnitude bins may be affected by biased lines
of sight.

In this paper we present an analysis of \hf~images obtained in the
NICMOS parallel program in 1997. In section 2 we describe the data
and our reduction methods. In section 3 we present the
galaxy number counts; in section 4 we discuss a
quantitative morphological classification of the galaxies and
compare the result to optical surveys.

\section{Archival Data and Data Reduction}

The parallel program consisted of ``JHK'' imaging using cameras 1 and 2
(hereafter NIC1 and NIC2) at their best focus until November 1997, and 
$J_{1.1}$~and \hf~imaging in Camera 3 (NIC3), as well as some grism
spectroscopy, thereafter.  \hf~data were taken in individual exposures
of 256s or 1024s for NIC2 and 256 or 576 seconds for NIC3.  We
selected 160 NIC2 pointings at galactic latitudes above 20\deg~that
are unaffected by foreground stars or the primary target and have no
instrumental oddities or other impediments (a total area of 15.2
square arcminutes). We also examined the first 30 NIC3 fields (20.2
square arcminutes).  An independent analysis of the NIC3 data is
presented by Yan \et (1998).

NICMOS parallel data are taken in ``MultiAccum'' mode, in which the
array is read out nondestructively many times during each exposure.
The resulting data cube contains multiple measurements of the
accumulated flux at increasing integration times. The count rate for
each pixel can be determined from the slope of a line fit to these
data (after dark subtraction). This procedure allows cosmic ray
rejection by identifying breaks in the slope.

NIC3 data were reduced using the CALNICA IRAF routines (Bushouse,
1997), using a model (noiseless) cube of multiaccum dark data
constructed by STScI and a skyflat derived from the median of all NIC3
\hf~observations available at the time.  NIC2 data were reduced
independently of the CALNIC routines which were still under
development at the time.  The major difference between our routines
and CALNICA is that we fit the slope of the multiaccum data before
flatfielding.  There was insufficient sky background
in the NIC2 data to construct a viable skyflat and the best
inflight pointed flat was used instead.

Individual reduced frames were registered for each field. For NIC3
data, we used the centroid of many objects to calculate the precise
offsets between dithered frames.  Registration was performed with
bilinear interpolation for fractional pixel shifts.  During
registration, a second cosmic ray rejection was performed using a
sigma-clipping algorithm. Many NIC2 fields contained no
bright objects for registration. In that case, we used the astrometry
information from the telescope pointing to calculate fractional pixel shifts.

Objects were identified using the SExtractor software (Bertin \&
Arnouts 1996), admitting no object with fewer than 6 pixels.
Identifications were carefully checked by eye to avoid false
detections from cosmic rays or noise at the vignetted edges of NIC3
frames. Photometry was performed in elliptical
apertures set at 2.5 times the Kron ``first moment'' radius (Kron
1980). 
We measured 1$\sigma$~limiting magnitudes by taking the standard
deviation of aperture photometry performed on several hundred randomly
selected positions throughout the frame.  In a typical photometric
aperture (3.5 pixel radius) we measured galaxies with H$_{1.6}$=22.8
for NIC2 (and 22.9 for NIC3) at the 3$\sigma$~level in 1024 second
exposures.  Limits are expressed in magnitudes on the Vega system. The
offset to the AB magnitude system (Oke 1974) is $+1.3$ for the
\hf~filter. The deepest fields reach H$_{1.6}\simeq 24.3$~in NIC2 and
24.0 in NIC3. In the 160 NIC2 fields, 142 extended objects were
detected. In the 30 NIC3 fields examined, more than 500 objects were
detected.

\section{Galaxy Counts}

We examine the galaxy content of the central ``overlap'' regions of
the registered images, that is the regions with full exposure time.  We
estimate completeness with simulations based on recovery of mock
objects inserted into real coadded images.  We find $>90$\%
completeness at the 5$\sigma$~level and 60\% completeness at
3$\sigma$.  We have included a small ($\sim 10\%$) correction to our
completeness estimate to account for reduced sensitivity to larger
objects (see Bershady \et~1998).
For objects bright enough for a reliable estimate of the full-width
half-maximum (based on the SExtractor algorithm), we separate
unresolved stars from the galaxies. In a typical field, this limit is
H$_{1.6}\sim 22.5$.  We then assume that all fainter objects are
galaxies, since there should be little contamination from faint stars
in these fields.  A similar assumption was made by Djorgovski \et
(1995) in making the deep K-band galaxy counts.

Figure 1 compares the $H_{1.6}-$band counts to the deepest
ground-based K-band counts, using a model prediction of their median
color (the {\it ncmod} program, Gardner 1998) to translate the counts
from the K-filter to the NICMOS \hf~filter.  For example, {\it ncmod}
predicts \hf$-K=0.91$ at \hf=23.  As has been seen in K-band data
(Bershady et al. 1998, Moustakas \et 1997), the $H_{1.6}$~galaxy
counts continue to rise at the faintest levels.  We note that the NIC3
counts appear less smooth than individual ground-based surveys or the
NIC2 counts.  While this effect could be real, it may be the result of
non-Poissonian errors (e.g.  field to field variations,
completeness,etc.)  Our future analysis of the entire NIC3 sample will
fully explore this question.  Our results show that NICMOS has already
reached H$_{1.6}\simeq 24.3$ at the 5$\sigma$~completeness limit,
using the most conservative error estimates. In the longest future
pointings, NICMOS parallel data will go deeper than current
ground-based IR number counts, and will accumulate more area.

\section{Morphology}

We take advantage of the high resolution of NICMOS images to 
evaluate galaxy morphology using the classification algorithm
of Abraham et al. (1996a).  The procedure uses 
the central concentration ($C$) and 
asymmetry ($A$), to separate E/S0's, spirals, and irregulars.
Ellipticals have strong central concentration, and low asymmetry.
Spirals show a considerable range of asymmetry and lower central
concentration.  Irregulars, peculiars and mergers lie in the
``exclusion zone'' of asymmetric and diffuse objects, where no
ellipticals or spirals are expected to fall.  

The asymmetry is determined by rotating the galaxy image about its center,
subtracting the original and measuring the ratio of flux in the
absolute value of the difference to that in the original. 
The central concentration compares the flux in an
aperture containing the inner third of the object as defined by the
second order moments to the total flux (Abraham \et 1994).
Both $C$ and $A$ are very sensitive to the size and center of the
aperture containing the object. Abraham \et defined the center of the
object to be the brightest pixel in the object after convolution with a
1 pixel wide gaussian. They defined the object aperture to be an
isophot enclosed by the pixels 1.5$\sigma$~above the sky. We have used
the same algorithm to determine $C$ and $A$, but we use the object
centroid as the center, and we use an elliptical aperture determined by
SExtractor, also based on 1.5$\sigma$~sensitivity.

To calibrate our classification routines, we have used a reference
sample of 81 ground-based images of E through Scd galaxies obtained by
Frei \et (1996) from Lowell observatory. This is the same comparison
set used by Abraham et al.  Comparing the values of $C$ and $A$ we
measure for this sample to the Abraham \et values, we obtain similar
results with a scatter roughly equal to the errors, $\sigma_{A} \sim
\sigma_{C} \sim \pm 0.05$.  We also obtain consistent $C$ and $A$
values for artificially redshifted versions of the Frei sample.
Abraham \et applied a complex artificial redshifting scheme, applying
a K-correction pixel by pixel. We have instead concentrated on
understanding the effects of the pixel scale.  We find that, as
expected, galaxies can appear more symmetric at lower resolution,
while their central concentration is almost unchanged, thereby
preventing a spurious increase in the number of irregular galaxies
when classified at NIC2 resolution.  To confirm this expectation, we
artificially redshifted the Frei sample to z = 0.3, 0.5, and 0.7 with
the NICMOS plate scale.  We continue to see a clean separation of
morphological types agreeing with the unredshifted type.  At higher z,
some galaxies cross the Elliptical/Spiral boundary in about equal
measure in each direction, but we see no suggestion that changing the
sampling or signal-to-noise of galaxies moves them into the
``irregular/peculiar'' part of the plane.

We automatically classified all galaxies with $\simgt$10$\sigma$, the
cutoff at which our ``eyeball'' classifications could be trusted
(H$_{1.6}<23$ in the deepest fields and 1 to 2 magnitudes brighter in
the shallowest).  Figure 2 shows the morphological classification of
all such objects in our NIC2 number counts sample (106 galaxies
total).  We find that based on our visual identification, the
Spiral/Elliptical difference could be identified mostly using central
concentration, while Spiral/Irregular separation relies heavily on
asymmetry.  However, no fewer ``misidentifications'' would be achieved
using a vertical cut in the C/A plane, so we retain the divisions of
Abraham et al., as borne out in our simulations.

We find few irregular galaxies (17\% of the total sample and
23\% of galaxies fainter than \hf=21). By comparison, the MDS survey
found that 21\% of the galaxies at $I_{814}<22$~are irregular, while
in the HDF the percentage rises to 30--40\% at 22$<I_{814}<25$ (Abraham \et
1996b). Our \hf=20--23 limits correspond to $I_{814}$=22.1--24.5 for a
mixture of elliptical and spiral galaxies, or $I_{814}$=21.3--24 for
late type spirals and irregulars.  This quantitative result is
confirmed by visual inspection; no ``train wrecks'' are seen (see Figure 3).

This fraction of faint, infrared-selected irregular galaxies appears
small compared to the number found in the MDS and HDF.  There are two
possible explanations for this effect: 1) We are seeing the same
galaxies but they are more regular in the NIR, or 2) we do not see all
of the irregulars because they are too blue.  Put another way, are we
seeing different stars or different galaxies?

To distinguish between these cases, we compare our results to a model
translation of the $I_{814}$-band counts (Gardner 1998).  Figure 4 shows
number counts segregated by morphological type.
For comparison, the number counts by type from
Abraham \et (1996b) and Glazebrook \et (1995) are shifted to the 
\hf~filter and plotted.
There is a clear difference in relative depth
between the optical and IR data for different morphological
types.  For the redder ellipticals, the NIC2 counts reach as
deep as the $I_{814}$~data, but for the blue irregulars the 
\hf~data reach only a magnitude shallower than the $I_{814}$~counts.
We see good agreement between the \hf~and $I_{814}$~number counts in
each morphological bin. In particular, though there are slightly fewer
irregular galaxies at \hf$<20$~in the \hf-band, they are in excellent
agreement at fainter magnitudes (the difference is likely due to the
small numbers of galaxies being counted).  Clearly, if we were seeing
a strong morphological K-correction, this difference would not be the
case. Such an effect would require that \hf~observations find less
irregulars at faint magnitudes than predicted based on
$I_{814}$~observations.

This result implies that the irregular galaxies observed in the MDS
are at intermediate redshifts ($z \sim 0.5$). At that redshift, the
observed $I_{814}$-band is approximately the rest-frame V band, while
the observed $H_{1.6}$-band is the restframe J-band. If the irregular
galaxies were at high redshift ($z>2$), where optical observations
sample the rest-frame ultraviolet, then there would be a large
morphological difference in the observed near infrared.  For example,
many Lyman Break Galaxies in the HDF are seen to have highly disturbed
morphologies (see Lowenthal \et 1996), which have been suggested by
O'Connell (1997) to be similar to the UV morphology of typical
spirals. This is consistent with the results of Glazebrook \et (1998),
who obtained spectra of 31 galaxies from the MDS irregular galaxy
sample and found them to be at typical redshifts of 0.5, though they
note that the irregulars are not necessarily a single uniform
population.

The relatively small number of irregulars in our sample is due to
their blue color. Glazebrook \et (1998) find irregular galaxies at
$I_{814}<22$~ have (observed) I-K colors of 2--3, with some red
outliers. Since the bulk of the irregulars are at I$>24$, that would
imply infrared magnitudes on the order of H$\simgt$22. This prediction
is consistent with our data, since we see some irregulars (possibly
the red outliers) but a lower percentage than was seen in the
$I_{814}$-band data. 

Could we see the high-redshift population in the NICMOS data? Typical
Lyman Break Galaxies have IR magnitudes of K=21--22 (Steidel \et
1996), and may have $22\simlt H_{1.6} \simlt 23$~and therefore be just
at the edge of our ability to classify in the current data (\hf$<23$).
Future NICMOS observations should be able to detect such galaxies,
though much of that work will be done using NIC3 with its coarser
plate scale, making morphological classification more difficult.

In summary, we have examined NICMOS parallel imaging data. We find
good agreement between the \hf~number counts and those in the deepest
ground based surveys.  We have morphologically classified the galaxies
in the NIC2 data following the quantitative classification scheme of
Abraham \et (1996a). We see evidence for the same population of faint
blue irregular galaxies observed in the MDS and HDF data.  We argue
that that the irregulars (at $H_{1.6}\sim$22) are at z$<1$.

\acknowledgments

We thank the members of the Space Telescope Imaging Spectrograph
Investigation Definition Team (STIS IDT) for their encouragement of
this project. We thank Jennifer Sandoval for assisting with obtaining
the data. We thank R. Abraham and K. Glazebrook for providing ASCII
tabulation of their morphologically segregated number counts.  We also
thank P. MacCarthy, L. Yan and L. Storri-Lombardi for useful
discussions.  We thank the referee, Matthew Bershady, for many helpful
comments which improved the manuscript.  This work was supported by
the STIS IDT through the National Optical Astronomical Observatories
and by the Goddard Space Flight Center.

\references
\reference{} Abraham, R.G., Valdes, F. Yee, H.K.C., \& van den Bergh, S., 
1994, \apj, 432, 75

\reference{} Abraham,R.G., van den Bergh,S., Glazebrook,K., Ellis,R.S., 
Santiago,B.X., 
Surma,P., \& Griffiths,R.E., 1996a, ApJS, 107, 1

\reference{} Abraham, R.G., Tanvir, N.R., Santiago, B.X., Ellis, R.S., Glazebrook, 
E., \&
van den Bergh, S. 1996b, MNRAS, 279, L47

\reference{} Abraham, R.G., 1998, in preparation.

\reference{} Bershady, M.A., Lowenthal, J.D., \& Koo, D.C., 1998, \apj in press

\reference{} Bertin, E. \& Arnouts, S., \aaps, 1996, 117, 393

\reference{} Bushouse, H., 1997, ``The NICMOS CALNICA and CALNICB Pipelines'' in
The 1997 HST Calibration Workshop, eds S. Casertano et al. 

\reference{} Djorgovski, S. \et, 1995, ApJL, 438, L13

\reference{} Driver, S.P., Windhorst, R.A., Ostrander, E.J., Keel, W.C., 
Griffiths, R.E., \& Ratnatunga, K.U., 1995, \apj, 449, 23

\reference{} Driver, S.P., Fernandez-Soto, A., Couch, W.J., Odewahn, S.C.,
Windhorst,R.A., Phillipps,S., Lanzetta,K., \& Yahil, A., 1998, ApJL, 496, L93

\reference{} Ellis, R.S., 1997, ARAA, 35

\reference{} Frei, Z., Guhathakurta, P., \& Gunn, J.E., 1996, \aj, 111, 174

\reference{} Gardner, J.P., 1998, PASP, 110, 291

\reference{} Gardner, J.P., Cowie, L.L., Wainscoat, R.J., 1993, \apjl 415, 9

\reference{} Glazebrook, K., Ellis, R., Santiago, B., \& Griffiths, R., 1995,
\mnras 275, L19

\reference{} Glazebrook, K, Abraham, R., Santiago, B., Ellis, R., \& Griffiths, R., 
1998, MNRAS in press

\reference{} Griffiths, R. E., et al.\ 1994, \apj, 435, L19

\reference{} Holfeltz, S.T., MacKenty, J.W., Sparks, W.B., \& Axon, D., 1998, 
BAAS 191.3.09

\reference{} Howell, S.B., 1989, PASP, 101, 616

\reference{} Hutchings, J.B., 1995, \aj 109, 928

\reference{} Kron, R. G. 1980, \apjs, 43, 305

\reference{} Loveday, J., Peterson, B.A., Efstathiou, G., Maddox, S.J., 1992,
 ApJ, 390, 338

\reference{} Lowenthal, J.D. \et, 1997, \apj, 481, 673

\reference{} Moustakas, L.A., Davis, M., Graham, J.R., Silk, J., Peterson, B.A., \&
Yoshi, Y., 1997, ApJ, 475, 445

\reference{} O'Connell, R. W., 1997, in 
The Ultraviolet Universe at Low and High Redshift, p.11, eds. Waller, W.H.

\reference{} Oke, J.B. 1974, \apjs, 27, 21

\reference{} Stanford, S.A., Eisenhardt, P.R.M., \& Dickinson, M., 1995, \apj, 450, 
512

\reference{} Steidel, C.C., Giavalisco, M., Pettini, M., Dickinson, M.,
\& Adelberger, K.L., 1996, ApJ Letters 462, L17

\reference{} Thompson, D., Djorgovski, S, \& Trauger, J., 1995, \aj, 110, 963

\reference{} Thompson, R. I., Rieke, M., Schneider, G., Hines, D.
C., \& Corbin, M., R. 1998, \apj, 492, L95

\reference{} Tyson, A., 1988, \aj 96, 1

\reference{} Williams et al 1996, \aj, 112, 1335

\reference{} Yan, L., McCarthy, P. J., Storrie-Lombardi, L. J., \&
Weymann, R. J. 1998, \apj, in press

\reference{} Yoshii, Y. \& Takahara, F., 1988, \apj, 326, 1

\clearpage

\figcaption[]{ Number counts for the \hf~filter. The blue circles show
  NIC2 and the darker blue sqaures show NIC3.  Arrows indicate $1\sigma$~upper
  limits.  The red + symbols indicated the ground-based K-band data (as
  compiled by Gardner, 1998 and Bershady et al., 1998) translated to
  the NICMOS \hf~filter. }

\figcaption{Morphological classifications of galaxies in the NIC2
  fields.  Classifications by eye are indicated by the symbols
  (stylized asterisks for irregulars, ellipses for E/S0, and stylized
  spirals for later than S0).  The data are plotted in the plane of
  central concentration (C) vs. asymmetry (A).  The lines indicate the
  division of morphological types according to Abraham et al.  (1996).
  }

\figcaption[]{A montage of classifiable galaxies from the NIC2 images.
  From top to bottom, galaxies are arranged in bins of ellipticals,
  spirals and irregulars according to the automated classification.
  Within each bin galaxies are arranged in descending order of
  apparent magnitude.  Each image is 3.1\arcs $\times$ 3.1\arcs.
  Objects identified differently by eye (see figure two) are indicated
  by the letter E,S, or I. }

\figcaption[]{Number Counts by morphological type.  The filled circles
  are the number counts for galaxies that can be morphologically
  classified.  The open squares indicate number counts of Abraham \et
  (1996b) and Glazebrook \et (1995) shifted to the \hf~filter. }

\clearpage

\begin{figure}[h]
\parbox{6in}{\epsfxsize=6in \epsfbox{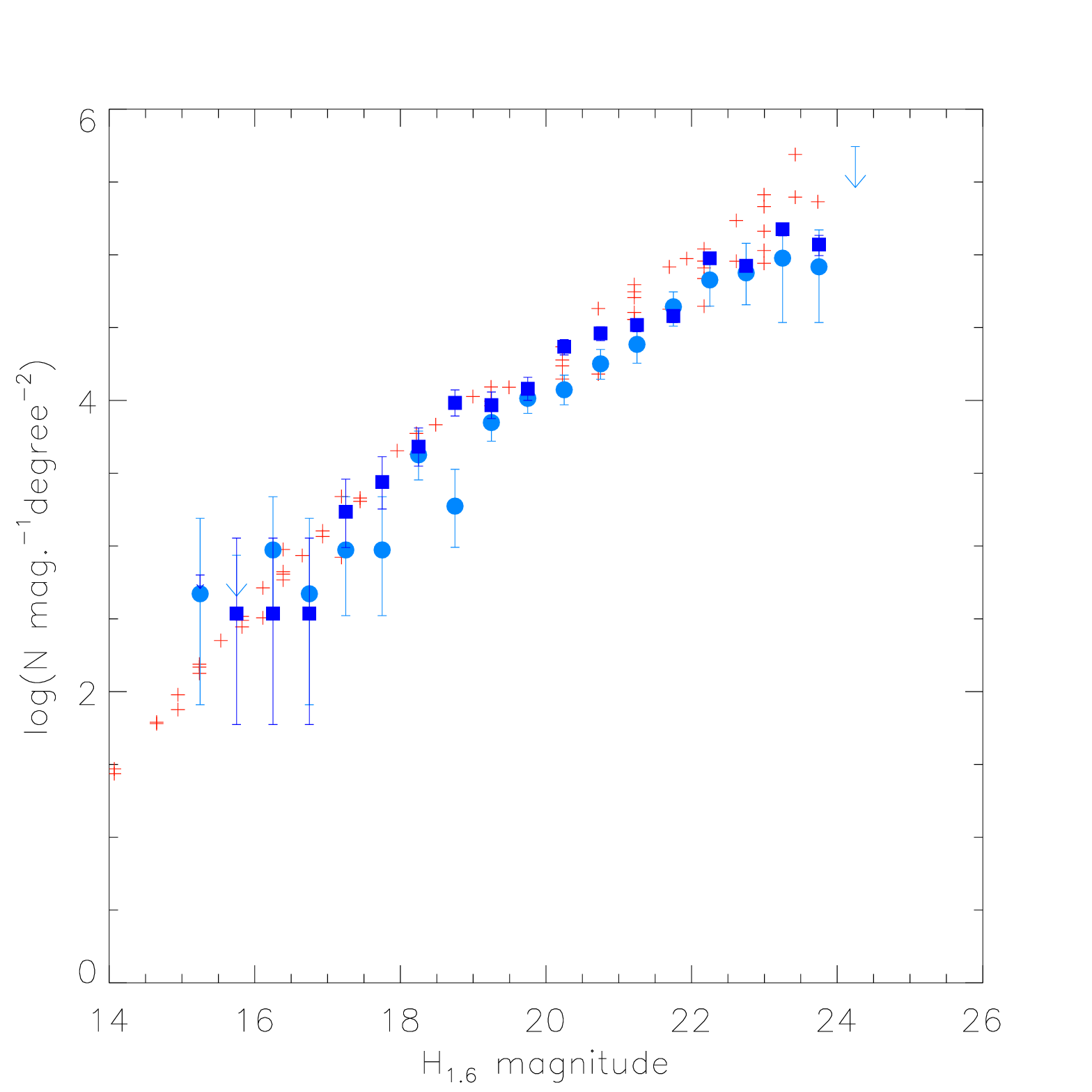}}
\end{figure}

\clearpage

\begin{figure}[h]
\parbox{6in}{\epsfxsize=6in \epsfbox{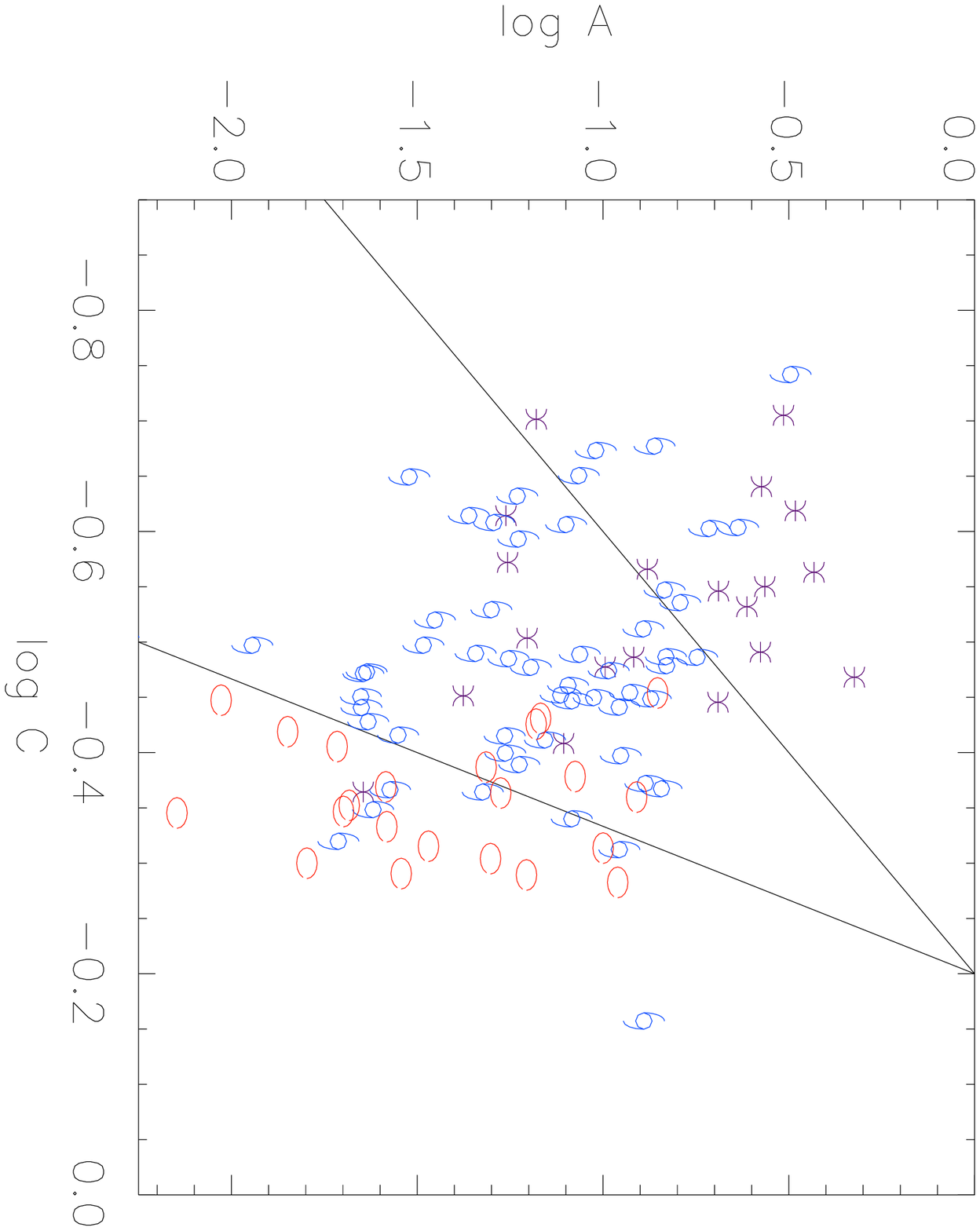}}
\end{figure}

\clearpage

\begin{figure}[h]
\parbox{8in}{\epsfysize=9in \epsfbox{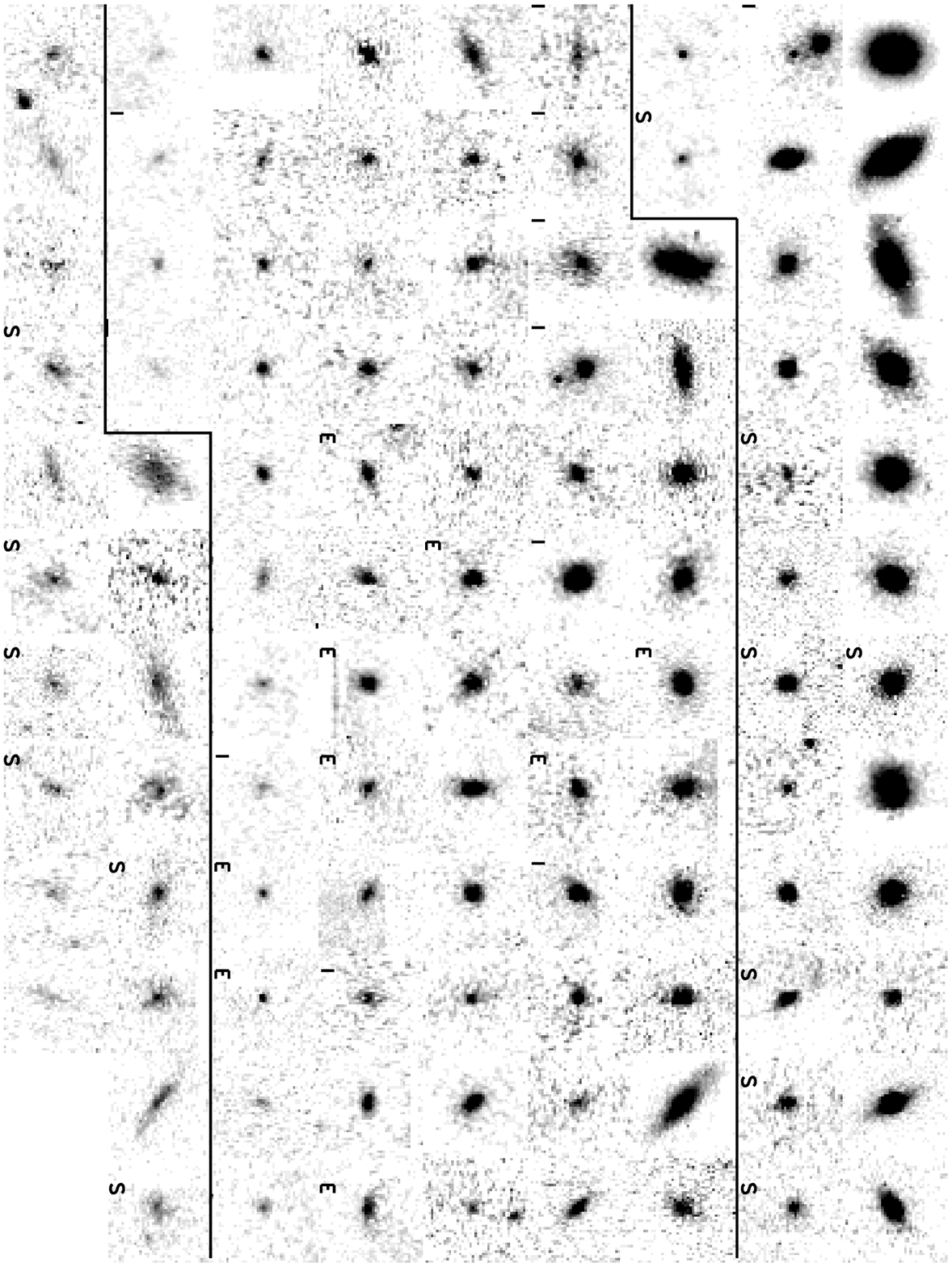}}
\end{figure}

\clearpage

\begin{figure}[h]
\parbox{6in}{\epsfxsize=6in \epsfbox{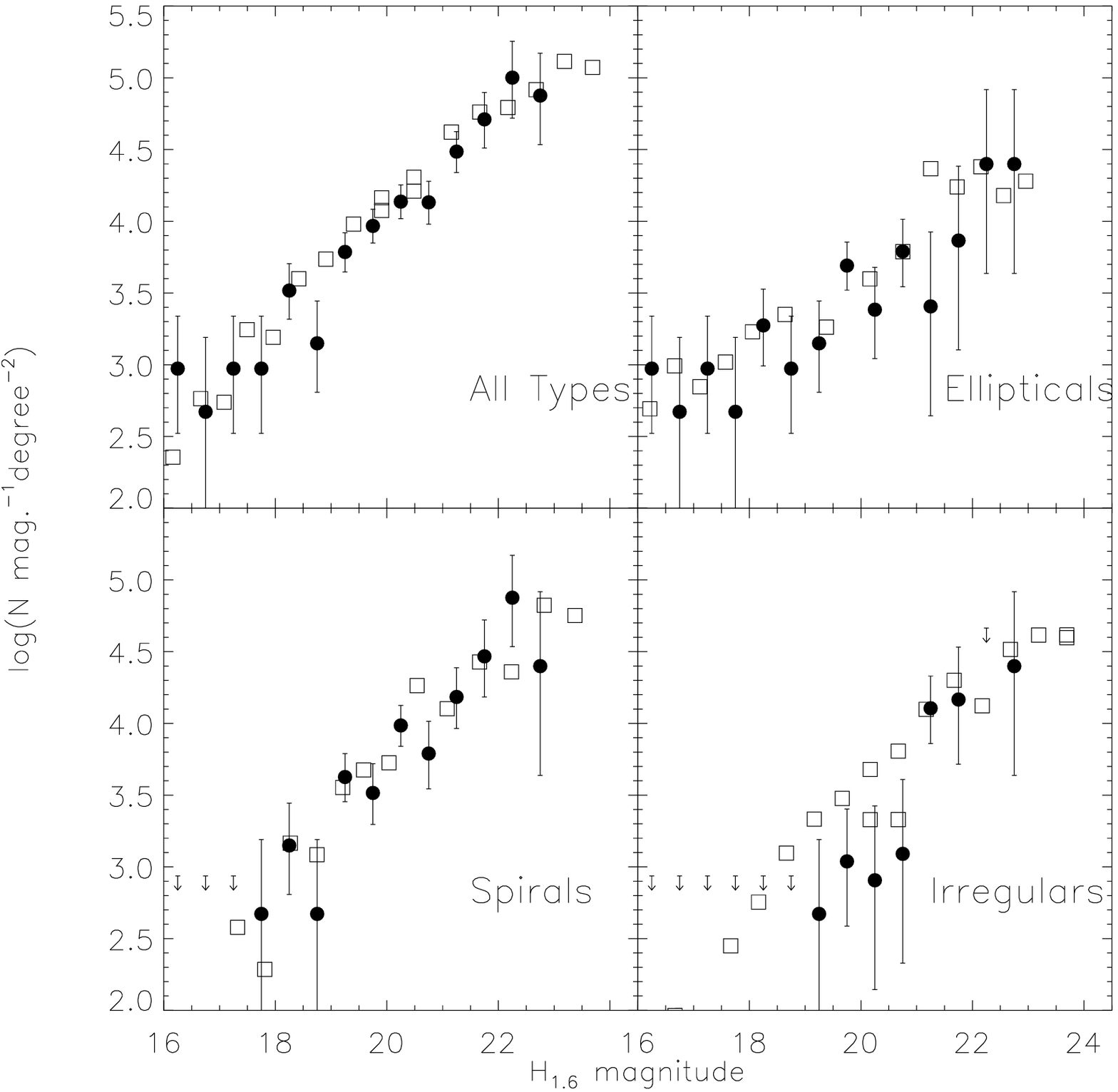}}

\end{figure}

\end{document}